 \documentclass[12pt,preprint]{aastex}

\usepackage{graphicx}
\usepackage{amsmath}
\def \apj {ApJ}

\def \apjl {ApJ}

\def \aap {A\&A}

\newcommand{\citeN}[1]{\citeauthor{#1} (\citeyear{#1})}
\newcommand{\citeNP}[1]{\citeauthor{#1} \citeyear{#1}}

\shortauthors{Socas-Navarro}
\shorttitle{Spectral Analysis based on Feature Extraction}

\begin{document}

%
%\title{Optimizing the Spatial Grid in the Solution of Inverse Problems}
\title{Feature Extraction Techniques for the Analysis of Spectral Polarization
  Profiles} 

\author{H. Socas-Navarro}
   	\affil{High Altitude Observatory, NCAR\thanks{The National Center
	for Atmospheric Research (NCAR) is sponsored by the National Science
	Foundation.}, 3450 Mitchell Lane, Boulder, CO 80307-3000, USA}
	\email{navarro@ucar.edu}

\date{}%

\begin{abstract}
This paper introduces a novel feature extraction technique for the
analysis of spectral line Stokes profiles. The procedure is based on the use
of an auto-associative artificial neural network containing non-linear hidden
layers. The neural network extracts a small subset of parameters from the
profiles (features), from which it is then able to reconstruct the original
profile. This new approach is compared to two other procedures that have been
proposed in previous works, namely principal component analysis and 
Hermitian function expansions. Depending on the target application, each one
of these three techniques has some advantages and disadvantages, which are
discussed here.
\end{abstract}


\section{Introduction}
\label{sec:intro}

A new breed of techniques for the diagnostics of solar magnetic fields has
been emerging during the last few years. The fundamental idea that inspired
them is the recognition that the observable polarization (Stokes) line
profiles exhibit conspicuous {\it patterns} of features associated with the 
underlying physical conditions in the atmosphere where they form. This
has recently motivated a considerable interest in the field of pattern
recognition. Potential applications to solar magnetic field diagnostics
have been explored in a number of recent papers by \citeN{RLAT+00};
\citeN{SNLAL01}; \citeN{CS01}; \citeN{SNSA02}; \citeN{SLA02}; \citeN{SN03};
\citeN{dTILA03}; \citeN{CLAT+03}.

Pattern recognition techniques were originally developed in the fields of
artificial intelligence and machine vision, but have found a vast
number of applications over a broad spectrum of disciplines (see, e.g., the
review by \citeNP{RG03}). Solar physicists have relied on least-squares
fitting procedures to invert the observed spectra since the late 70s,
although this only became a generalized practice in the early 90s (for more
details see the reviews by \citeNP{SN01a}; \citeNP{dTI03}). The
least-squares approach is still as useful as ever and is not likely to be
replaced by the newer pattern recognition in the near future. The reason is
simply that one can obtain much more detailed information from a given
set of profiles by using a least-squares inversion. It also offers more
flexibility in the physical model used for the inversion. Research on new
inversion techniques is intended to develop alternative 
algorithms that will {\it complement}, and not {\it replace} previous
least-square procedures. 

Nevertheless, it is justified to devote significant efforts towards the
development of alternative inversion algorithms. Without going into much
detail here (see, e.g., \citeNP{SN02} for a more thorough discussion), it 
should be noted that the
next generation spectro-polarimeters (both ground-based and space-borne) will 
deliver enormous dataflows that simply cannot be analyzed with traditional
inversion codes. The need for larger fields of view, higher spatial
resolution and better time cadence is driving new instrumental
developments. These, in turn, demand new analysis techniques that are fast,
ideally operating in real-time, and robust, in the sense of not requiring any
human supervision. Other important considerations that motivate research on
pattern recognition are related to the need for establishing long-term
databases of observations (properly inverted), or the desire to obtain
real-time maps of the solar magnetic field at the observing site.

As an intermediate step in this process, some authors have sought a suitable
set of features in the Stokes profiles of spectral lines and an appropriate
algorithm to extract such features. Feature extraction is of great potential
interest, e.g. for data compression (which would help circumvent limited
spacecraft telemetry), dimensionality reduction, profile classification,
inversion pre-processing, etc. The most promising approaches developed so far
seem to be the use of principal component analysis (PCA), introduced by
\citeN{RLAT+00}, and the expansion in Hermitian functions (EHF) of
\citeN{dTILA03}. In this paper I present a new method which makes use of
artificial neural networks (ANNs), and compare it to PCA and EHF in terms of
performance and potential usefulness. 

The organization of the present paper is as
follows. Section~\ref{sec:expansions} below reviews the PCA and EHF methods
and discusses some important ideas that are relevant for our purposes
here. Section~\ref{sec:ANN} gives a detailed introduction to ANNs and
puts forward the concept of auto-associative ANNs, which is the base for the
feature extraction algorithm.
%\footnote{
%A toolbox of computer routines (both in FORTRAN
%and IDL languages) for building and training ANNs is provided with the CD-ROM
%format of this publication. 
%}
The comparison among the three methods is presented in
\S\ref{sec:performance}. Finally, the most important conclusions are
discussed in \S\ref{sec:conc}.

\section{Profile expansion in basis functions}
\label{sec:expansions}

Mathematically, the problem considered in this paper can be summarized as
follows. Given a set of one or more Stokes profiles $p^k(\lambda_i)$ (where
$k=1,\dots,4$ denotes the Stokes parameter and $\lambda_i$ are the
$N_{\lambda}$ wavelengths of the spectral samples), we wish to extract a
suitable set of $M$ parameters ${\bf c}$ (features) that contain as much
information from the profiles as possible: 
\begin{equation}
{\bf c}={\bf H}({\bf p}) \, ,
\end{equation}
with ${\bf H}$ being a $M$-dimensional function that characterizes the
feature-extraction technique. We can then reconstruct the original profile
approximately by doing the inverse transformation:
\begin{equation}
{\bf p} \simeq {\bf p'} = {\bf R}({\bf c}) \, .
\end{equation}

Three different strategies are explored in the next sections. The first two,
PCA and EHF, are series expansions in terms of orthonormal basis functions.
%simple linear transformations, which means that the $c_j$
%are linear combinations of the $p^k(\lambda_i)$. 
The third method, AANN, is
somewhat more complex and allows for non-linear relations between the
features and the profiles.

\subsection{Principal component analysis}
\label{sec:PCA}

PCA has been used in a broad variety of fields (e.g., \citeNP{RG03}) and was
recently introduced in 
the radiative transfer literature by \citeN{RLAT+00}. At its core, PCA is
simply a series expansion in terms of some orthonormal basis functions. If
$p(\lambda_i)$ is a discretized spectral profile, we have:

\begin{equation}
\label{eq:PCA}
p(\lambda_i)=\sum_{j=1}^{N} c_j b^j(\lambda_i) \, ,
\end{equation}
where $N$ is the number of wavelength samples, $b^j(\lambda_i)$ are the basis
functions (sometimes referred to as eigenprofiles or eigenvectors), and $c_j$
are the coefficients of the expansion. These coefficients are sometimes
referred to as features or eigenfeatures. 

The basic premise that makes PCA so special is that the basis is cleverly
chosen for each particular problem that one is dealing with. Instead of using
a predefined set of basis functions, PCA seeks an {\it optimum} (in the sense
defined below) set of eigenprofiles $b^j(\lambda)$. The first obvious
condition that the $b^j(\lambda)$ must meet in order to form a basis of the
profile space is orthonormality:

\begin{equation}
\label{eq:PCA2}
\sum_{i=1}^N b^j(\lambda_i) b^k(\lambda_i) = \delta_{jk} \, ,
\end{equation}
where $\delta_{jk}$ is the Kronecker delta. This condition implies that the
expansion coefficients $c_j$ can be determined simply by the following scalar
product:

\begin{equation}
\label{eq:PCA3}
c_j = \sum_{i=1}^N p(\lambda_i) b^j(\lambda_i) \, .
\end{equation}

Equation~(\ref{eq:PCA}) is exact because the sum extends to as many as $N$
expansion terms. However, the whole point of feature extraction is precisely
to reduce the dimensionality of the problem. This goal is achieved by
truncating the expansion at a given order $M$ ($M < N$), retaining only
those terms that are deemed relevant. Consider a statistical sample of $N_p$
profiles, denoted by ${\bf p^j}$ (with $j=1,\dots,N_p$). The individual
components of a vector ${\bf p^j}$ are the observable spectral samples
($p_i^j = p^j(\lambda_i)$). Let us approximate these profiles by:

\begin{equation}
\label{eq:PCA4}
{\bf p^i} \simeq \sum_{j=1}^{M}  c^i_j {\bf b^j} + 
\sum_{j=M+1}^N e_j {\bf b^j} \, ,
\end{equation}
where ${\bf e}$ is a constant vector. When the expansion is truncated at
order $M$, the total quadratic error accumulated over the entire sample is:
 
\begin{equation}
\label{eq:PCA5}
E_M=\sum_{i=1}^{N_p} \sum_{j=M+1}^N (c^i_j - e_j)^2 \, .
\end{equation}

Notice that, so far, we have not particularized to PCA or given an explicit
form for the ${\bf b^j}$. The discussion above applies to any expansion in
basis functions. The distinctive feature of PCA is that it provides a basis
that minimizes the truncation error for any order
$M$. Minimization of $E_M$ with respect to $e_j$ straighforwardly leads to:

\begin{equation}
\label{eq:ej}
e_j = {1 \over N_p} \sum_{i=1}^{N_p} c^i_j \, .
\end{equation}
The truncation error $E_M$ can also be minimized with respect to the basis
functions ${\bf b^j}$. One then obtains:

\begin{equation}
\label{eq:eigen}
{\bf C b^j} = \lambda_j {\bf b^j} \, ,
\end{equation}
where ${\bf C}$ is the covariance matrix, defined as:

\begin{equation}
\label{eq:covmatrix}
{\bf C} = \sum_{i=1}^{N_p} ({\bf p^i - \bar p}) ({\bf p^i - \bar p})^T \, .
\end{equation}
In Eq~(\ref{eq:covmatrix}) above, ${\bf \bar p}$ is the average profile over
the statistical sample and $T$ denotes the usual matrix transposition
operation.

Equation~(\ref{eq:eigen}) shows that the PCA basis functions are simply the
eigenvectors of the covariance matrix. Hence the denominations of
eigenfunctions or eigenprofiles that they sometimes receive. The PCA
expansion procedure can be summarized as follows. First, one picks a suitable
database of profiles that are deemed representative of the physical problem
being tackled. This database should ideally span a broad range of profiles,
covering all the possible realizations that we might expect to find in the
observations. The next step is to construct the covariance matrix ${\bf C}$
by using Eq~(\ref{eq:covmatrix}), and solve the corresponding eigenvalue
problem (Eq~[\ref{eq:eigen}]). The eigenvectors of ${\bf C}$ are a basis of
orthonormal profiles ${\bf b^j}$ ($j=1,\dots,N$) with some interesting
properties. In particular, it provides the {\it optimum} linear coordinate
transformation for our 
problem (assuming that the original database is a good representation), in
the sense that the information preserved upon truncation at any order $M$ is
maximum. 
%Notice that the database may be constructed from either synthetic or
%observed profiles. Both approaches have advantages and disadvantages, as
%discussed in \S\ref{sec:conceptual} below.

\subsection{Hermite functions}
\label{sec:EHF}

The Hermite functions constitute an orthonormal basis of ${\cal L}^2$, the
space of integrable functions. They can be calculated as the product of a
Gaussian by the Hermite polynomials (with some normalization constants):

\begin{equation}
\label{eq:Herm1}
H^n(x)= {1 \over \sqrt{n! 2^n \sqrt{\pi}}} \exp(-x^2) {\cal H}^n(x) \, .
\end{equation}

The Hermite polynomials that appear in Eq~(\ref{eq:Herm1}) are given by the
following recursive relations:

\begin{eqnarray}
\label{eq:Herm2}
{\cal H}^0 & = & 1  \nonumber \\
{\cal H}^1 & = & 2 x  \nonumber \\
{\cal H}^n & = & 2(n-1)  x H_{n-1} - 2 H_{n-2}  \, .
\end{eqnarray}

The $H^n(x)$ defined by in Eqs~(\ref{eq:Herm1}) and~(\ref{eq:Herm2}) are
orthonormal. As with PCA above, any given profile may be expanded as linear
combination of this basis:

\begin{equation}
\label{eq:Herm3}
p(x)=\sum_{j=1}^{N} c_j H^j(x) \, .
\end{equation}
Obviously, the optimal expansion is attained when $x$ is the wavelength
relative to line center, and normalized to the width of the line.

\section{Artificial Neural Networks}
\label{sec:ANN}

\begin{figure*}
\epsscale{.8}
\plotone{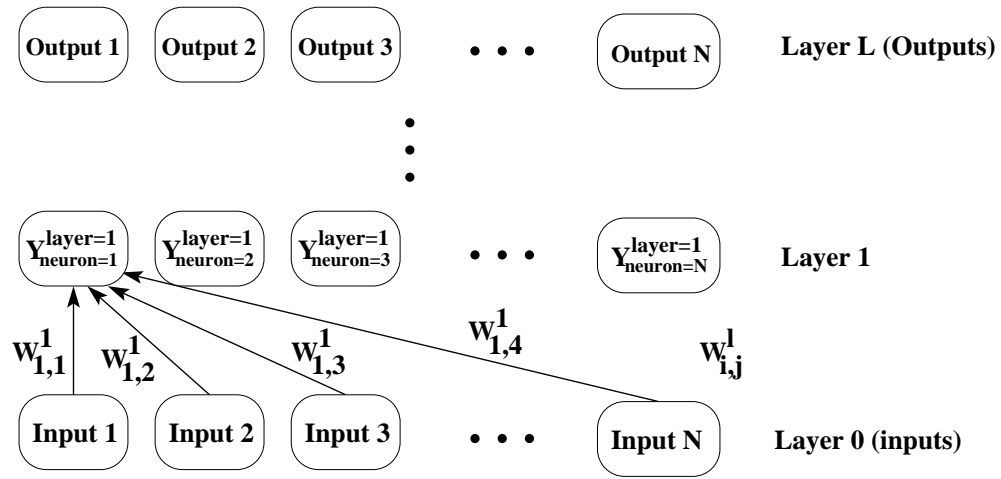}
\caption{
\label{fig:ANN}
Schematic representation of a multi-layer perceptron.
}
\end{figure*}

Before going into details on the particular physical problem of spectral line
analysis, a general introduction to ANNs is probably in order.
An ANN is an abstract structure that mimics to some extent the functioning of
a biological neural system. Consider a number of memory cells (neurons),
each one having the ability to store a datum. The neurons are inter-connected
({\it synaptic connections}) in such a way that the number they store can be
modified according to the values stored in other neighboring neurons that is
is connected to (see Fig~\ref{fig:ANN}).  

ANN investigations usually employ some simplifying restrictions on the
structure of neurons and synaptic connections.
%In order to make
%the problem tractable, some simplifying restrictions are usually applied. 
One
of them is to consider {\it forward-propagating} networks, in which a signal
propagation direction is defined. The synaptic connections seen by a neuron
can be inward- or outward-directed, depending on whether it is used to
receive or send a signal from or to a neighbor. Some neurons are taken as
inputs in which the signal originates and have no inward connections. The
neurons connected to the inputs are then 
modified according to the values in the input neurons. Such modifications in
turn trigger a new set of neurons to alter their values, and the process
continues until the output neurons (which have no further outwards
connections) are reached. Forward-propagating networks are not allowed to
have loopback connections. 

In order to fix ideas let us consider a special type of forward-propagating
ANN, namely the {\it multi-layer perceptron} (MLP). This
is perhaps the most widely used ANN geometry. A MLP is
arranged in successive layers, as sketched in Fig~\ref{fig:ANN}. Every neuron 
in a given layer is connected with all neurons in the previous layer. No
conexions are allowed between neurons in the same layer, or between layers
that are not successive. 
\clearpage
\clearpage
Let us discuss in some more detail the actual process of signal
propagation. We start with a set of inputs $x_i$ (with
$i=1,\dots,N$). Denoting by $Y^l_n$ the datum stored in neuron $n$ belonging
to layer $l$, and defining $l=0$ to be the input layer, we have that
\begin{equation}
Y^0_i = x_i \, .
\end{equation}

The basic equation for signal propagation in a MLP is:

\begin{equation}
\label{ANNeq}
Y^l_n = f \left ( \sum_{j=1}^N W^l_{n,j} Y^{l-1}_j + \beta^l_n \right ) \, ,
\end{equation}
with $l=1,\dots,N_{layers}$ and $n=1,\dots,N$. The $W^l_{n,j}$ and
$\beta^l_n$ are the synaptic weights and biases, respectively. They are
parameters of the network that define its behavior. The function $f$ is the
so-called {\it activation function}. A MLP is said to be linear when
$f(x)=x$. In this case the value in a given neuron is simply a linear
combination of the data in its neighbors. For many interesting applications,
however, non-linear activation functions are chosen. One of the most widely
employed functions is the hyperbolic tangent, which shall be used in this
paper as well.

From a mathematical point of view, a forward-propagating ANN may be viewed as a
multi-dimensional non-linear function ${\bf F}$ that transforms a vector of
inputs 
$x_i$ (with $i=1,\dots,N$) onto a vector of outputs $o_j$ (with
$j=1,\dots,M$). The input and output spaces are not necessarily of the same
dimension ($N \ne M$). Formally:

\begin{equation}
\label{eq:ANN}
{\bf o} = {\bf F} ( {\bf x} ) \, .
\end{equation}

The idea is to use ${\bf F}$ as an approximation to the physical model under
consideration. What is distinctive about ANNs is their ability to
``learn''. This means that the actual form of ${\bf F}$ is altered during a
{\it training process}, making it a better approximation. 

The usual procedure to train a MLP is by generating a set of training data,
consisting of known inputs $x^t_i$ and outputs (sometimes referred to as
targets) $o^t_i$. The training inputs are propagated through the network,
producing some outputs $o_i$, which in general differ from the targets by an
error $\delta_i=o_i - o^t_i$. One then modifies the network parameters
$W^l_{ij}$ and $\beta^l_i$ to minimize the sum of quadratic errors. Several
algorithms exist in the literature to perform this task. Perhaps the most
widely used is the {\it back-propagation} algorithm (\citeNP{RHW86}), which
has been employed 
for the calculations in the present paper. Computer routines in Fortran~90 to
create and train MLPs are available from the author upon request.
%The reader can find 
%IDL and Fortran90 routines to create and train MLPs in the electronic version
%of this journal.

\subsection{Auto-associative networks}

\begin{figure*}
\epsscale{.8}
\plotone{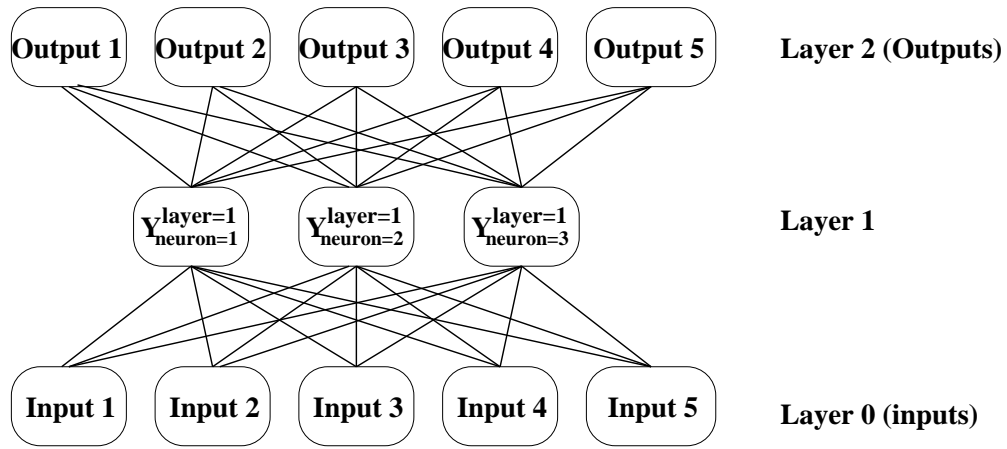}
\caption{
\label{fig:AANN}
Schematic representation of an auto-associative network. This particular
network has one hidden layer with 3 neurons. The input and output layers have
5 neurons.
}
\end{figure*}

Dimensionality reduction (and feature extraction) can be achieved by using
{\it auto-associative} ANNs (hereafter AANNs). An AANN is trained with
targets that are equal to the inputs of the training set ($o^t_i =
x^t_i$). This would not be very useful without another distinctive feature of
AANNs, namely a reduced number of neurons in one (or more) of the hidden
layers (the ``bottleneck'' layer). The resulting structure for a simple AANN
is shown in Fig~\ref{fig:AANN}. 

\begin{figure*}
\epsscale{.8}
\plotone{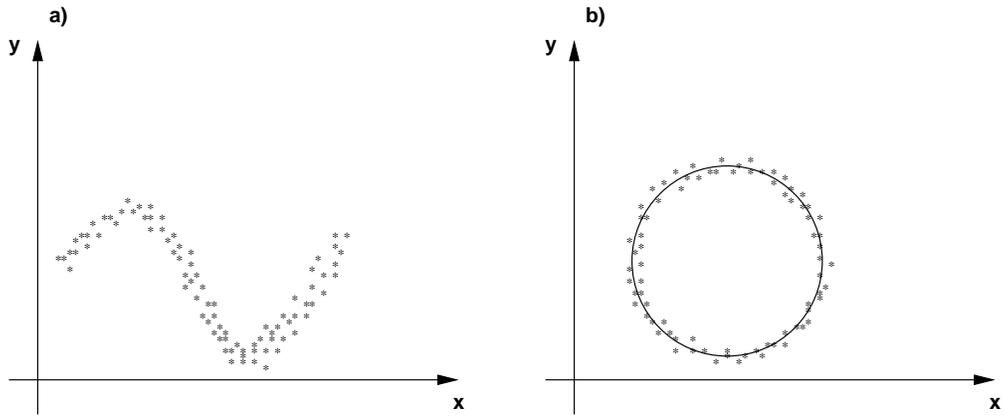}
\caption{
\label{fig:circle}
Two simple problems that illustrate the advantages of non-linear coordinate
transformations.
}
\end{figure*}

When the AANN in Fig~\ref{fig:AANN} is trained with $o^t_i = x^t_i$, it needs
to find a way to compress the input data into a smaller subset of numbers (in
the particular example of the figure, going from 5 to 3) from which the input
profile can be later reconstructed. When the AANN has only one hidden layer
(as the one in the figure) and the activation function is linear, it can be
shown that the AANN training has one global solution (\citeNP{BK88};
\citeNP{BH89}). This solution is precisely the same that one wold obtain with
a PCA decomposition. In other words, a properly trained linear AANN with one
hidden layer is able to extract PCA coefficients from an input profile and
then reconstruct the original profile from these coefficients.
\clearpage
\clearpage
Using AANNs to do PCA decomposition is not very useful. There are simpler
methods that are guaranteed to work and do not require a cumbersome,
time-consuming 
training process. However, there may be instances in which it is convenient
to use multi-layer 
non-linear AANNs, which are effectively some form of non-linear
generalization of PCA. Extreme examples are the academic
problems represented in Fig~\ref{fig:circle}. 
\clearpage
\clearpage
In the $(x,y)$ coordinate system, one needs two values to entirely determine
the position of the points in the figure. A PCA analysis of the dataset would
result in another orthogonal reference system $(\mu,\nu)$, which is simply a 
rotation of $(x,y)$ thus requiring also two numbers for each
datapoint. Obviously, the optimal coordinate choice for these problems would
be a system that has one coordinate $\eta$ along the curve and another $\rho$
perpendicular to it. In this new sytem, specifying $\eta$ alone practially
determines the location of the datapoints. 

We can say in this case that the ``intrinsic dimensionality'' of the problem
is $D_{dim}=1$, which is smaller than that of the ``measurement'' space
$D_{meas}=2$. A similar reasoning may be applied to the analysis of spectral
Stokes profiles. Typical observations of a line profile consist of
$\sim$100 spectral samples times 4~Stokes parameters. However, the intrinsic
dimensionality of the physical problem is usually much smaller. For instance,
a simple Milne-Eddington model is specified by $\simeq$10~atmospheric
parameters. 

The mapping from $(x,y)$ to $(\eta,\rho)$ in the examples above is non-linear
and therefore cannot be picked up by the PCA decomposition. However, a
properly trained AANN may be able to perform this mapping. This is an
interesting property with potential interest for spectral line diagnostics.

\section{Comparison}
\label{sec:performance}

In this section I present some results from application of the three methods
(PCA, EHF and AANN) to the analysis of simulated solar profiles. A word
of caution is in order, since these results depend on the particular
conditions of the problem under study, such as the set of profiles
used for the simulation, the model atmosphere considered, etc. Therefore, the 
performances reported here are to be taken as merely orientative.

For a sake of computational simplicity, the simulations have been carried out
with Milne-Eddington models that have fixed thermodynamical parameters,
corresponding to typical solar conditions. The
magnetic field vector varies randomly, with the field strength ranging
between 0 and 4000~G, and both the inclination and azimuth between 0
and 180 degrees.\footnote{
Notice that it is not necessary to simulate azimuths greater than 180 degrees
because of the well-known ambiguity of the Zeeman-induced profiles.}
The spectral lines simulated are the well-known pair of \ion{Fe}{1} lines at
6302~\AA . Random noise at the level of $10^{-3}$ was added to all
profiles. 

Several sets of 5000 profiles were computed from random models. One of such
sets was used to calculate the PCA $b_j$ eigenprofiles (see the discussion
preceding Eq~[\ref{eq:eigen}]). The rest were used to train an AANN with
a total of 4 non-linear layers. The spectral ranges corresponding to the two
telluric lines in this region have been removed from the input profiles, but
the network is still required to provide the full profile on output.
The ``bottleneck'' layer of the AANN contains 10 neurons, whereas the input
and output layers have 320 each. The reason for chosing 10 neurons in the
bottleneck layer is that this is the intrinsic dimensionality of the profiles
(given here by the number of free parameters in the Milne-Eddington
model atmosphere). 
%It can be shown that a MLP with a sufficiently large
%number of neurons is able to approximate any continuous multi-dimensional
%mapping to an arbitrary precision (e.g., \citeNP{J90}; \citeNP{BL91}). This
%property of MLPs makes it possible for 

\begin{figure*}
\epsscale{.8}
\plotone{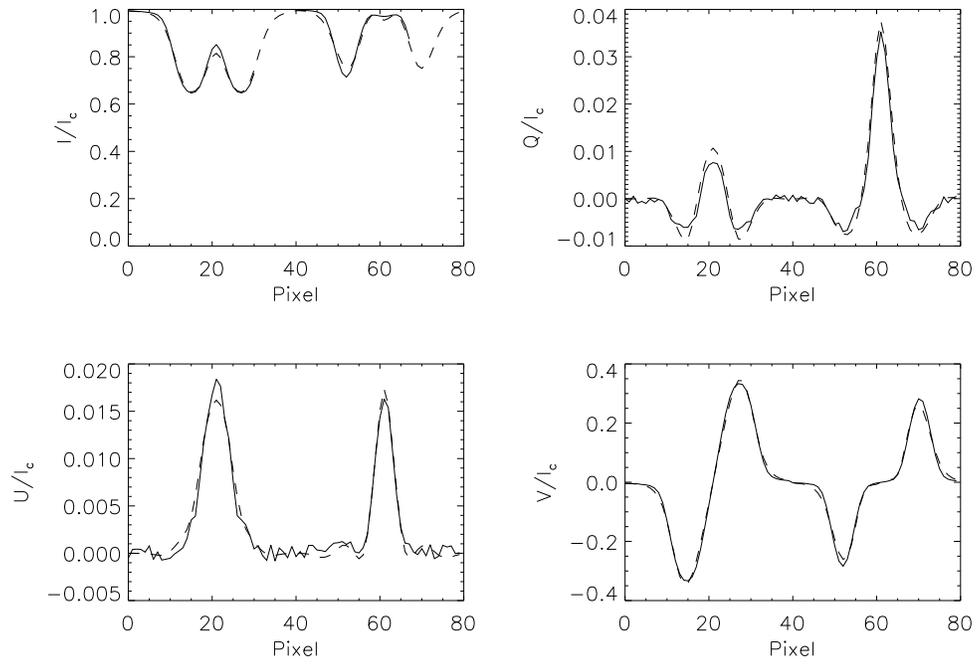}
\caption{
\label{fig:profs}
Simulated Stokes profiles (solid) normalized to the continuum intensity
($I_c$), and fit from the AANN (dashed). 
}
\end{figure*}

After a given training 
batch has been processed the ANN is presented with a set of 500 test profiles
(not included in the training) to measure its performance. The training
process was stopped when two successive batches yield no significant
improvement in this test. The total CPU time for the training process was
over 30 hours running on a 2.4~GHz Intel 
Pentium~4 processor. Fig~\ref{fig:profs} shows the final
reconstruction accuracy of the AANN for a sample set of Stokes profiles. 
\clearpage
\clearpage
A reference sample of 500 spectra was analyzed and reconstructed using PCA,
the EHF and the AANN. For each spectrum, the difference
between the simulated and the reconstructed profiles is quantified as:
\begin{equation}
\label{chisq}
\chi^2 = {1 \over N_{data}} \sum_{i=1}^{N_{data}} { (I^{sim}_i -
  I^{rec}_i ) ^2 \over   \sigma_i^2 } \, ,
\end{equation}
where $N_{data}$ is the number of spectral samples multiplied by the
number of Stokes parameters considered, $I^{sim}$ is the reference profile,
$I^{rec}$ 
is the reconstruction and $\sigma_i$ is the measurement noise. Using this
definition, the error in the reconstruction shown in
Fig~\ref{fig:profs} is $\chi^2=10.27$.

\begin{figure*}
\plotone{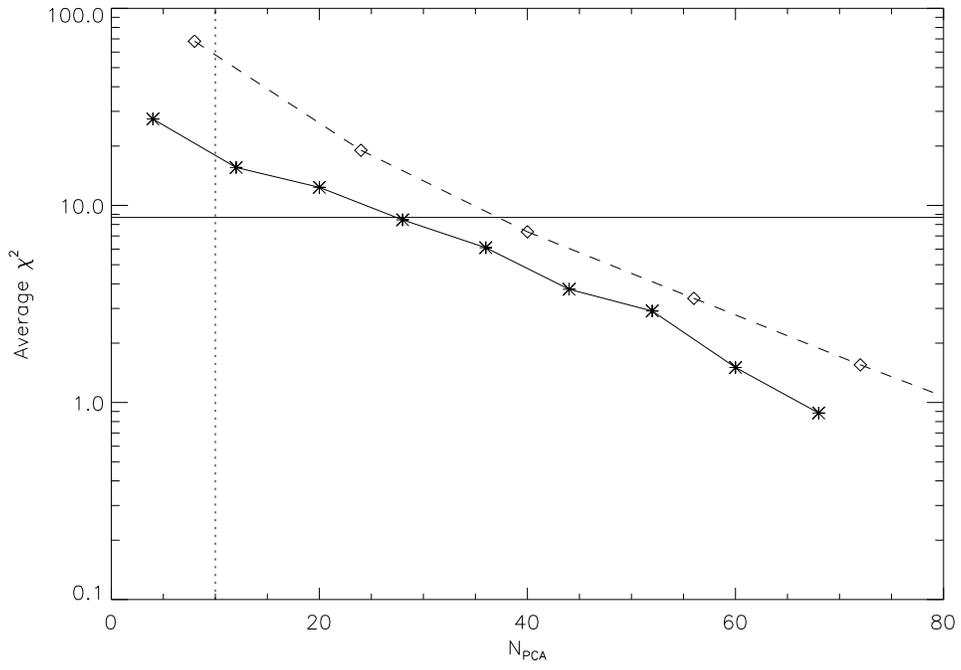}
\caption{Average $\chi^2$ for a reference sample of 500 random
  profiles. Solid: PCA. Dashed: EHF. The horizontal and
  vertical lines mark the reconstruction error of the AANN and the number of
  neurons in the bottleneck layer, respectively.
\label{fig:chi2}
}
\end{figure*}

Figure~\ref{fig:chi2} shows the $\chi^2$, averaged over the reference
sample, as a function of the coefficients used 
in the PCA and EHF. In the
case of the AANN the number of parameters is fixed and determined by the
structure of the network. The vertical line marks the number of neurons in
the bottleneck layer.

Not surprisingly, the EHF exhibits the
lowest accuracy. However, this method still has merit as it is the only
one that has analytical basis functions. Moreover, it does not require any
``a priori'' computations, unlike the AANN (which needs training) or PCA
(which needs a database). The AANN works well when high
accuracy is not required and the model atmospheres are relatively simple (see
the reconstruction in Fig~\ref{fig:profs}). Unfortunately, the computational
expense increases rapidly for more complex models or higher accuracy
demands, rendering this technique impractical for some problems. Both PCA
and the EHF ultimately converge to zero error when a
sufficiently large number of expansion coefficients is employed. This is 
probably a desirable feature for some applications, depending on the
reconstruction error tolerance. Once it has been trained, the AANN
forward-propagation is very
fast because it only requires the calculation of $Y^l_n$ for each neuron
using Eq~(\ref{ANNeq}). Notice that we only need to propagate the signal from
the inputs to the bottleneck layer (in the case of feature extraction), or
from the bottleneck to the outputs (in the case of profile
reconstruction). In terms of computational requirements, none of the three
techniques considered here presents a clear advantage over the others.

\section{Conclusions}
\label{sec:conc}

The development of new instrumentations and the need to process ever
increasing dataflows is turning the attention of solar physicists to pattern
recognition techniques (see references in \S\ref{sec:intro}).
The present work can be considered a follow up to the papers by
\citeN{RLAT+00} and \citeN{dTILA03}, introducing a new technique for feature
extraction and comparing it to the proposed PCA and EHF proposed in those
papers.  

The AANN expansion is based on a MLP which is trained to reproduce
the profile supplied in the input neurons. However, one (or more) of the MLP
hidden layers has fewer neurons than the inputs/outputs layers. Upon
successful training, the AANN is able to extract a small set of features from
a profile. These features are numbers stored in the bottleneck layer, from
which the entire profile can be reconstructed. This works effectively as a
decomposition/expansion scheme in which the number of coefficients is set by
the number of neurons in the bottleneck layer.

The comparison among the three feature extraction techniques does not reveal
a ``winner''. Instead, each method has its own advantages and
disadvantages. Depending on the particular problem it may be more interesting
to use one or another. 

Ultimately, the goal is to use the expansion coefficients from whichever
method one might choose for the implementation of some inversion
procedure. The most straightforward way to do this would be a look-up table,
as \citeN{SNLAL01} did for PCA. Other alternatives include neural
networks or support vector machines, which could benefit from the
dimensionality reduction offered by these techniques as a pre-processing
layer. 
\clearpage
%\bibliographystyle{../bib/apj}
%\bibliography{../bib/aamnem99,../bib/articulos}

\end{document}